\documentclass[11pt]{article}

\usepackage[final]{acl}

\usepackage{times}
\usepackage{latexsym}

\usepackage[T1]{fontenc}

\usepackage[utf8]{inputenc}

\usepackage{microtype}

\usepackage{inconsolata}

\usepackage{graphicx}

\usepackage{amsmath,amsfonts,bm}

\def\eqref#1{equation~\ref{#1}}

\def\1{\bm{1}}

\DeclareMathAlphabet{\mathsfit}{\encodingdefault}{\sfdefault}{m}{sl}
\SetMathAlphabet{\mathsfit}{bold}{\encodingdefault}{\sfdefault}{bx}{n}

\usepackage{hyperref}
\usepackage{url}
\usepackage{booktabs}
\usepackage{tabularx}
\usepackage{makecell}    
\usepackage{array}
\usepackage{caption}
\usepackage{enumitem}  
\usepackage{url}      
\usepackage{hyperref}
\usepackage{listings}
\usepackage{amsthm}
\usepackage{wrapfig}
\usepackage{float}

\usepackage{xcolor}

\usepackage{tcolorbox}
\theoremstyle{definition}
\newtheorem{definition}{Definition}
\usepackage{tcolorbox}
\tcbuselibrary{listings,breakable}
\newtcolorbox{promptbox}[1][]{%
  colback=gray!5!white,
  colframe=gray!80!black,
  fonttitle=\bfseries,
  title=Prompt,
  breakable,
  width=\linewidth,
  before=\noindent,
  left skip=0pt,
  right skip=0pt,
  boxrule=0.5pt,
  left=4pt,
  right=4pt,
  top=4pt,
  bottom=4pt,
  #1
}

\title{\textsc{Duet}: Joint Exploration of User–Item Profiles in Recommendation System}

\author{
\textbf{Yue Chen\textsuperscript{1,*}},
\textbf{Yifei Sun\textsuperscript{1,*}},
\textbf{Lu Wang\textsuperscript{2,$\dagger$}},
\textbf{Fangkai Yang\textsuperscript{2}},
\textbf{Pu Zhao\textsuperscript{2}},
\textbf{Minjie Hong\textsuperscript{3}},
\textbf{Yifei Dong\textsuperscript{4}},
\textbf{Minghua He\textsuperscript{2}},
\\
\textbf{Nan Hu\textsuperscript{2}},
\textbf{Jianjin Zhang\textsuperscript{2}},
\textbf{Zhiwei Dai\textsuperscript{2}},
\textbf{Yuefeng Zhan\textsuperscript{2}},
\textbf{Weihao Han\textsuperscript{2}},
\textbf{Hao Sun\textsuperscript{2}},
\\
\textbf{Qingwei Lin\textsuperscript{2}},
\textbf{Weiwei Deng\textsuperscript{2}},
\textbf{Feng Sun\textsuperscript{2}},
\textbf{Qi Zhang\textsuperscript{2}},
\textbf{Saravan Rajmohan\textsuperscript{2}},
\textbf{Dongmei Zhang\textsuperscript{2}}
\\
\\
\textsuperscript{1}Peking University 
\textsuperscript{2}Microsoft 
\textsuperscript{3}Zhejiang University 
\textsuperscript{4}KTH Royal Institute of Technology \\
\textsuperscript{*}Equal contribution
\textsuperscript{$\dagger$}Corresponding author
}

\newcommand{\ourmethod}{\textsc{DUET}~}
\begin{document}
\maketitle

\begin{abstract}
Traditional recommendation systems represent users and items as dense vectors and learn to align them in a shared latent space for relevance estimation. Recent LLM-based recommenders instead leverage natural-language representations that are easier to interpret and integrate with downstream reasoning modules. This paper studies how to construct effective \textit{textual profiles} for users and items, and how to align them for recommendation.
A central difficulty is that the best profile format is not known a priori: manually designed templates can be brittle and misaligned with task objectives. Moreover, generating user and item profiles independently may produce descriptions that are individually plausible yet semantically inconsistent for a specific user--item pair. We propose \textsc{Duet}, an interaction-aware profile generator that jointly produces user and item profiles conditioned on both user history and item evidence. \textsc{Duet} follows a three-stage procedure: it first turns raw histories and metadata into compact cues, then expands these cues into paired profile prompts and then generate profiles, and finally optimizes the generation policy with reinforcement learning using downstream recommendation performance as feedback. Experiments on three real-world datasets show that \textsc{Duet} consistently outperforms strong baselines, demonstrating the benefits of template-free profile exploration and joint user--item textual alignment. Project code page: \href{https://github.com/duet-review/duet_code}{https://github.com/duet-review/duet\_code}.
\end{abstract}

\section{Introduction}

Traditional recommendation systems represent users and items as dense vectors and learn to align them in a shared latent space for relevance estimation~\citep{covington2016deep, wu2023computational}. While effective, such embeddings are opaque: they offer limited interpretability and make it difficult to analyze why an item is recommended. Recent work therefore leverages large language models (LLMs) to introduce semantically rich, human-readable representations for recommendation~\citep{wang2025letting, zhang2024guided, bao2023tallrec,hong2025apoenhancingreasoningability, hong2025eager, hong2025generativereasoningrecommendationllms, wang2024eager}. A natural direction is to replace latent vectors with textual user and item profiles that can be inspected, edited, and reused by downstream components incorporated with LLMs.
However, existing LLM-based approaches remain limited in two important ways. First, directly prompting an LLM with raw user and item histories to obtain recommendations often yields noisy and incomplete signals, especially when histories are long, sparse, or heterogeneous~\citep{wang2025letting}. Second, profile-based methods typically rely on manually designed templates or handcrafted attributes, which requires substantial human engineering and constrains the representation space. More fundamentally, many approaches generate user profiles and item profiles independently, without modeling how user preferences and item semantics interact at decision time~\citep{yang2023palr, xi2024towards}.

\begin{figure}
    \centering
    \includegraphics[width=1\linewidth]{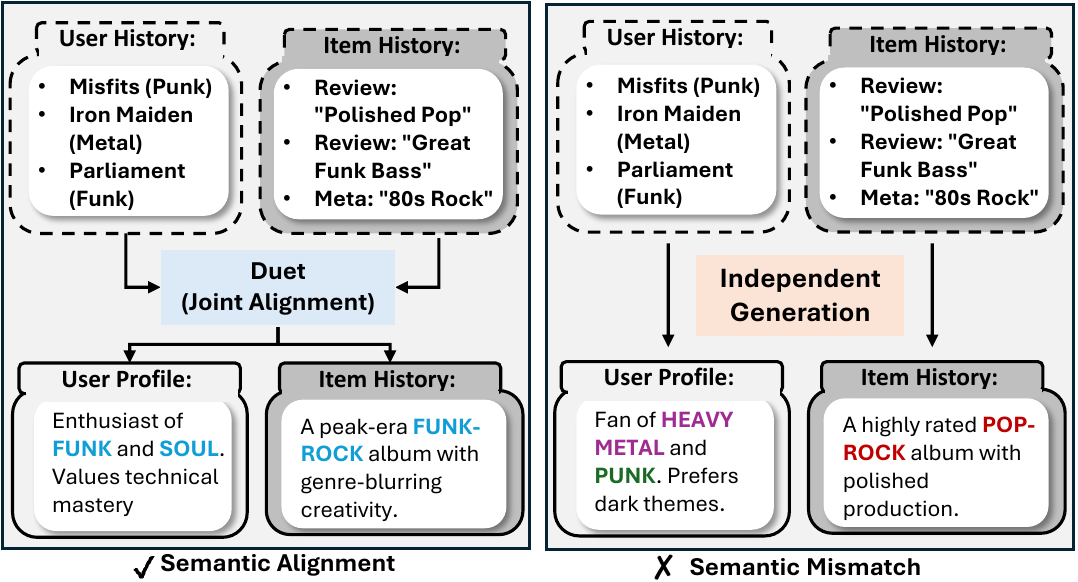}
    \caption{\textsc{Duet} aligns raw user and item data by transforming them into textual profiles within a shared semantic space. }
    \label{fig:teaser_2}
    \vspace{-4mm}
\end{figure}

To address these challenges, we propose \textsc{Duet}, a joint user--item profile generator that takes both user history and item history as input and produces a paired set of profiles for the interaction. Crucially, \textsc{Duet} does not require profile templates: it is trained with reinforcement learning using feedback from downstream recommendation performance, enabling it to explore and discover effective profile formats automatically. Figure~\ref{fig:teaser_2} illustrates why joint profiling matters. The user has listened to Misfits (punk), Iron Maiden (metal), and Parliament (funk), while the candidate album is described by reviews such as ``polished pop'' and ``great funk bass'' and a meta tag ``'80s rock.'' Considering both sides together, \textsc{Duet} can reconcile these signals into a compatible interpretation (e.g., highlighting the user's funk/soul affinity and the item's funk-rock character). In contrast, independently generated profiles may amplify different facets, summarizing the user as ``heavy metal/punk'' while summarizing the item as ``pop-rock'', which resulting in a semantically mismatched pair that obscures the true relevance signal.

\textsc{Duet} proceeds in three stages. First, raw histories and metadata are distilled into minimal \emph{cues} that capture compact but informative signals. Second, the model expands cues into richer profile prompts to generate textual profiles, allowing exploration over alternative profile structures and emphases. Third, the resulting profiles are consumed by downstream recommenders, and their task feedback is used to optimize the profile generation policy. By coupling both sides in a shared semantic space, \textsc{Duet} learns user profiles that reflect what kinds of items a user prefers and item profiles that reflect what kinds of users an item appeals to.

Our contributions are as follows:
\begin{itemize}[leftmargin=*,noitemsep,topsep=2pt]
    \item We represent users and items as natural-language profiles and align them in a shared semantic space, extending the classic vector-based alignment principle to interpretable textual representations.
    \item We introduce an exploration-based framework that starts from cue-based initialization, expands cues into candidate profiles, and \emph{jointly} optimizes user and item profiles with downstream recommendation feedback via reinforcement learning, avoiding rigid templates.
    \item Extensive experiments across multiple real-world datasets show that \textsc{Duet} consistently outperforms strong baselines, validating both joint profiling and feedback-driven profile optimization.
\end{itemize}

\section{Related Work}
\subsection{Profiles in Recommendation}
Early recommendation systems primarily relied on pre-defined profiles based on structured attributes, as seen in works like CRESDUP~\cite{chen2007content} and UPCSim~\cite{widiyaningtyas2021user}. While foundational, these methods were limited by their rigid, hand-engineered features. More recently, the advent of Large Language Models (LLMs) has enabled a shift towards generating profiles in natural language. Studies such as KAR~\cite{xi2024towards}, GPG~\cite{zhang2024guided}, PALR~\cite{yang2023palr}, and LettinGo~\cite{wang2025letting} leverage LLMs to create textual user profiles from behavioral data. A key limitation of these approaches is twofold: they rely on static or \textbf{pre-defined templates} for profile generation, and they often focus exclusively on user profiles, neglecting the rich, expressive information inherent in items and the complex dynamics of user-item interactions. In contrast, our work introduces a new paradigm where both user and item profiles are not fixed but are dynamically explored and optimized in a shared semantic space to directly align with recommendation performance.

\subsection{Reinforcement Learning for LLM-Based Recommendation Systems}
Reinforcement Learning (RL), particularly through techniques like RLHF, has become a core method for aligning LLMs with specific objectives. This approach has been adapted for recommendation tasks~\cite{wang2025letting,lin2025recr1,deng2025onerec}, but existing efforts often face key limitations. They frequently rely on offline reward models~\citep{jeong2023factual, chen2025warriormath, he-etal-2025-execoder, liu2025wedlmreconcilingdiffusionlanguage} that do not adapt in real-time to system feedback, a setup that risks issues like reward hacking~\citep{skalse2025rewardhacking}. Other methods~\citep{sun2024rlrf4rec,lu2024aligning} restrict themselves to offline preference tuning (e.g., DPO), which can easily overfit on static datasets. Our framework, \textsc{Duet}, overcomes these challenges by integrating RL into a closed-loop system where downstream recommendation performance serves as the real-time reward signal, allowing for the dynamic and interactive refinement of textual profiles.

\section{Method}
\begin{figure*}
    \centering
    \includegraphics[width=1\linewidth]{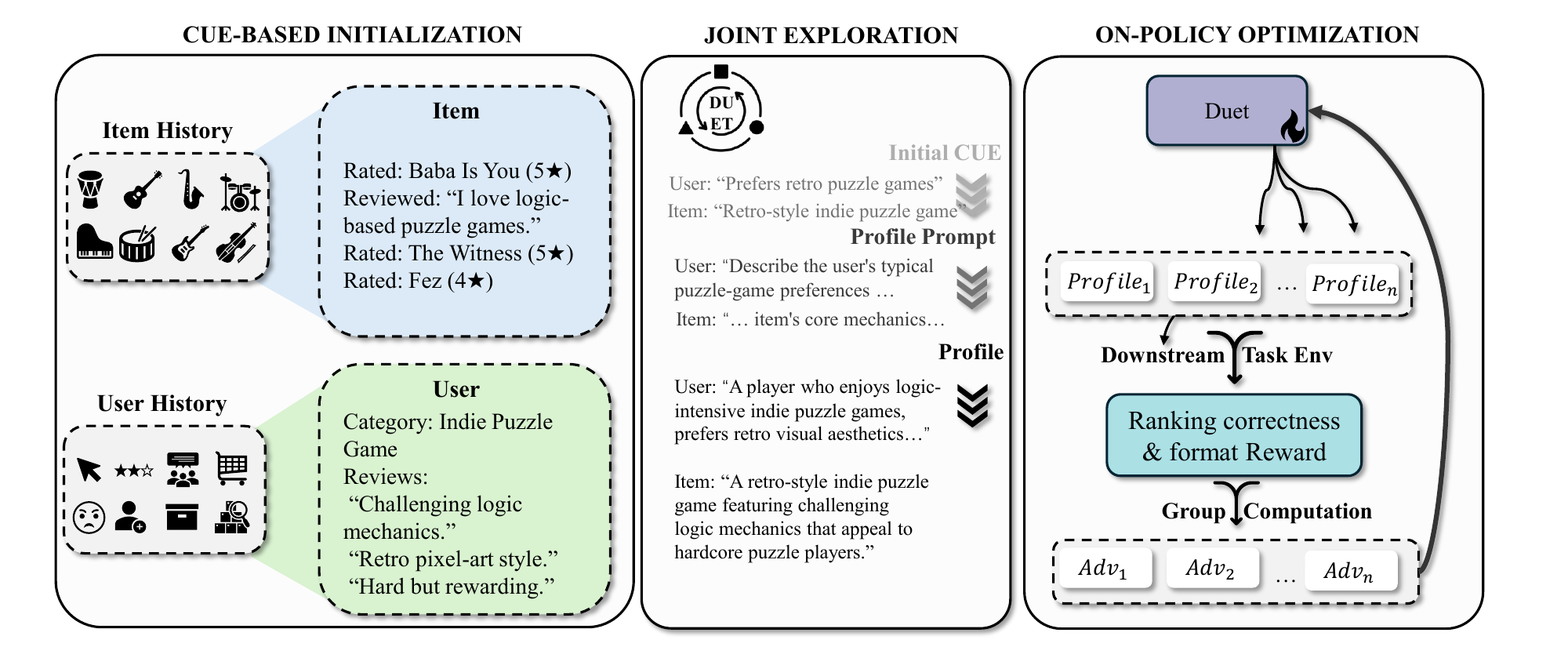}
    \caption{Overview of the \textsc{Duet} framework.}
    \label{fig:pipline}
\end{figure*}

\textsc{Duet} is a closed-loop framework that transforms raw user--item interaction histories into performance-aligned textual profiles through learned representation strategies. As shown in Figure~\ref{fig:pipline}, \textsc{Duet} consists of three stages. (1) \textbf{Cue-Based Initialization} distills interaction histories into concise evidence-based cues. (2) \textbf{Joint Exploration via Adaptive Profile Prompt Discovery} jointly explores user-item's profile prompts that define how user and item profiles should be written. (3) \textbf{Optimization via On-policy Exploration} jointly optimizes user and item profiles under downstream recommendation feedback.

All three stages are realized through a single pass input and output: cue extraction, self-prompt construction, and profile generation are produced in a single sequence-to-sequence generation pass at inference time, enabling efficient deployment.

\subsection{Problem Formulation}

We formulate profile generation in \textsc{Duet} as an \textbf{on-policy reinforcement learning} problem, motivated by the absence of any textual ground truth defining an optimal user or item profile. Profile quality is evaluated solely by its functional utility in a fixed recommendation environment.

\textsc{Duet} is modeled as a generative policy $\pi_\theta$ interacting with a frozen downstream recommender. For each user--item pair, the state is defined as $s=\{H_u,H_i\}$, where $H_u$ and $H_i$ denote the user and item interaction histories.
An action corresponds to a single-pass joint generation
\[
a=\{C_u,S_u,P_u,C_i,S_i,P_i\},
\]
where $(C_u,C_i)$ are cues distilled from history, $(S_u,S_i)$ are constructed user-item's profile prompts, and $(P_u,P_i)$ are the final textual profiles.

The policy $\pi_\theta(a\mid s)$ defines a joint distribution over the entire generation sequence. As an on-policy agent, \textsc{Duet} is optimized using rewards from its own sampled generations rather than by imitating fixed summaries.

\subsection{Cue-Based Initialization}

Raw user histories and item metadata are often noisy, redundant, and not directly suitable for profile construction. 
To address this, \textsc{Duet} introduces the concept of \emph{cues}: concise hypotheses that summarize minimal but informative aspects of users and items. 
These cues act as lightweight seeds, which are deliberately underspecified so that the system can subsequently explore richer profile formats.

\begin{definition}[Cue]
A \emph{cue} is a minimal textual hypothesis derived from historical data that highlights one potential aspect of a user’s preference or an item’s characteristic. 
Rather than aiming for completeness, cues capture partial but salient signals that serve as starting points for profile exploration.
\end{definition}

To extract cues automatically, the LLM is prompted to summarize minimal but informative aspects of user or item data. 
For example, given a user’s interaction history, the model is guided with instructions such as:

\begin{tcolorbox}[
    colback=gray!5!white,      
    colframe=gray!60!black,    
    title=Cue Extraction Prompt, 
    fonttitle=\bfseries,       
]
\emph{``From the history below, analyze the user's historical interactions to understand preferences, rating behavior, review sentiment or any other dimension. Keep the description concise and avoid full sentences.''}
\end{tcolorbox}

This lightweight guidance allows the LLM to map raw histories and metadata into compact textual cues. The detaild example of cue can be found in Appendix~\ref{app:cue}.

\subsection{Joint Exploration via Adaptive Profile Prompt Discovery}

After cue extraction, \textsc{Duet} does not directly summarize user interests or item attributes. Instead, it explores the space of \emph{profile construction strategies}—natural-language prompts that define the format, abstraction level, and attribute selection logic used to generate profiles. Exploration is therefore performed over how profiles should be constructed, rather than over superficial textual paraphrases.

For each user–item pair, \textsc{Duet} introduces an explicit intermediate variable, the \texttt{constructed\_prompt}, which serves as a discrete and interpretable profile prompt $S$. This profile prompt is a short natural-language instruction (e.g., \emph{``Describe the user's typical gaming preferences and engagement patterns''}) that specifies what aspects to describe and how to organize them, without containing the profile content itself. Conditioned on $S$, the model generates the final user and item profiles.

Exploration is driven by treating the profile prompt $S$ as a stochastic action sampled from the policy $\pi_\theta(S \mid \text{Cue})$. During training, the agent samples different profile prompt instructions, executes them to generate profile pairs $(P_u, P_i)$, and receives a reward based on downstream recommendation accuracy. The policy is optimized to reinforce profile prompts that consistently yield higher rewards, enabling the model to actively search for effective profile construction formats.

This process is realized as a unified generation pass,
\begin{equation}
    O = [\underbrace{\text{Cue}}_{\text{Context}} \rightarrow \underbrace{\text{profile prompt } S}_{\text{Profile Construction Prompt}} \rightarrow \underbrace{\text{Profile}}_{\text{Execution}}] ,
\end{equation}
where the profile prompt $S$ parameterizes the profile generator. At inference time, \textsc{Duet} collapses the learned profile prompt distribution and greedily executes the optimal profile prompt in a single forward pass, introducing no additional latency.

\paragraph{Example of profile prompt Exploration.} 

As illustrated in Figure~\ref{fig:prompt_flow}, \textsc{Duet} resolves this by identifying a shared context in a single forward pass. The policy extracts a coordinated \textbf{Cue} (e.g., ``prefers retro puzzle games'') and constructs a \textbf{profile prompt} (the \texttt{constructed\_prompt}) that instructs the model to highlight ``nostalgic aesthetics and strategic logic'' for the user, while simultaneously describing the item’s ``pixel-art graphics and intellectual difficulty'' to match. RL reinforces this shared semantic direction, suppressing irrelevant signals and forcing the final profiles to converge into a \textbf{shared semantic space} of nostalgia and logic, significantly improving recommendation accuracy.


\begin{figure}[!t]
\centering
\small
\setlength{\fboxsep}{4pt} 

\colorbox{gray!10}{%
\begin{minipage}[t]{0.98\linewidth}
\textbf{Single-Pass Input}\\
\textbf{Task:} Generate structured profiles for the user and item from historical data.\\
\textbf{Combined Context:} \texttt{[User History $H_u$]} + \texttt{[Item History $H_i$]} + \texttt{[Avg Ratings]}\\
\emph{``First, extract a concise \textbf{Cue} for both user and item. Second, based on the cues, construct a \textbf{Profile Prompt} that defines the description logic. Finally, generate the \textbf{Profile} guided by the constructed Profile Prompt.''}
\end{minipage}%
}

\vspace{2pt}
{\footnotesize $\downarrow$ \textbf{\textsc{Duet} Forward Pass} $\downarrow$}
\vspace{2pt}

\colorbox{green!5}{%
\begin{minipage}[t]{0.98\linewidth}
\textbf{Single-Pass Output}\\
\textbf{User}\\
\hspace*{1em}\textbf{Cue:} \emph{``prefers nostalgic logic-based challenges''}\\
\hspace*{1em}\textbf{Profile prompt:} \emph{``Focus on 1990s visual aesthetics and strategic depth.''}\\
\hspace*{1em}\textbf{Profile:} \emph{``A player who seeks retro-style visual charm paired with deep strategic reasoning.''}\\
\\
\textbf{Item}\\
\hspace*{1em}\textbf{Cue:} \emph{``retro-style indie puzzle with high difficulty''}\\
\hspace*{1em}\textbf{Profile prompt:} \emph{``Describe pixel-art graphics and intellectual difficulty to match logic preference.''}\\
\hspace*{1em}\textbf{Profile:} \emph{``A 2D experience featuring pixelated nostalgia and challenging mechanics that demand logical deduction.''}
\end{minipage}%
}

\caption{Single-pass generation in \textsc{Duet}: cue extraction, profile prompt (constructed prompt), and profile generation are produced in one pass for both user and item.}
\label{fig:prompt_flow}
\end{figure}

\subsection{Optimization via On-policy Exploration}

\paragraph{On-policy optimization.}
We train the profile generator $\pi_\theta$ in an on-policy manner against a frozen downstream model $f$, which serves as the environment critic. For each sampled user–item pair $(u,i)$, the policy generates $(P_u,P_i)$ and receives a scalar reward based on the prediction accuracy of $f(P_u,P_i)$. The policy parameters are updated to reinforce generations that lead to lower prediction error.

\paragraph{Continuous fractional reward.}
Using discrete integer ratings as rewards leads to sparse and unstable feedback, as near-miss predictions (e.g., predicting $4$ for a ground truth $5$) receive the same penalty as severe errors. To provide dense and informative feedback, we define a continuous fractional reward:
\begin{equation}
R_{\text{perf}}(u,i) = 1 - \frac{|y_{ui} - \hat{y}_{ui}|}{M},
\end{equation}
where $\hat{y}_{ui} = f(P_u,P_i)$ is the predicted relevance score and $M$ is the maximum rating gap (e.g., $M=4$ for a $1$–$5$ scale). This reward provides fine-grained gradients that encourage incremental improvements in recommendation accuracy.

We adopt Group Relative Policy Optimization (GRPO)~\cite{deepseekai2025} to optimize the profile generator under the above reward. The downstream recommender $f$ is frozen throughout training, which yields a stable  policy optimization setting and prevents reward drift or feedback-induced representation collapse.

\section{Experiment}

\subsection{Experimental Settings}\label{sec:exp_setting}

\noindent \textit{\textbf{Datasets.}}
Experiments were conducted on three widely used real-world datasets. \textbf{Amazon Music (Music)} and \textbf{Amazon Book (Book)} are derived from the Amazon Product dataset\footnote{\url{https://cseweb.ucsd.edu/~jmcauley/datasets/amazon/links.html}}, while \textbf{Yelp} is from the Yelp Open dataset\footnote{\url{https://business.yelp.com/data/resources/open-dataset/}}. All datasets include user reviews, ratings, and rich textual information. We used the full Amazon Music dataset, but only subsets (latest two months for Book and six for Yelp). Data was split by timestamp into training, validation, and test sets to prevent information leakage~\cite{Ji2023DataLeakage}.

\noindent \textit{\textbf{Evaluation Metrics}}
 
For each observed user–item interaction (i.e., a review record), we construct the evaluation instance based on the user’s interaction history strictly prior to the corresponding timestamp. Specifically, the user’s recent historical interactions before the current interaction are used to generate a \emph{user profile}, while historical reviews from other users are used to generate an \emph{item profile}. The downstream recommendation system then predicts the rating of the target item conditioned on two inputs: the generated user profile, and the generated item profile. An exception is the \textbf{10H} baseline, for which the downstream model directly consumes the raw recent interaction histories.

We evaluate performance using four widely adopted metrics~\cite{wang2025letting,fang2025reason4rec}: \textbf{Mean Absolute Error (MAE)}, \textbf{Root Mean Square Error (RMSE)}, \textbf{Accuracy}, and \textbf{F1 score}. 
In addition to rating prediction, we further evaluate the generated user and item profiles under a ranking setting based on downstream predicted rating scores. 
For each observed user--item interaction, we randomly sample nine items that the user has not interacted with to form a candidate set of ten items, which are ranked in descending order according to their predicted ratings. 
The ground-truth interacted item is treated as the only positive instance, and we adopt \textbf{NDCG@K} as the evaluation metric, with \textbf{K} set to \textbf{1}, \textbf{5}, and \textbf{10}, to assess how effectively the learned profiles support correct item ranking.

\noindent \textit{\textbf{Baselines}}
We compare our method with several representative baselines. \textbf{10H} directly uses the most recent interaction histories for prediction without constructing explicit profiles. \textbf{KAR}~\cite{xi2024towards} augments recommendation models with external reasoning knowledge about user preferences and factual knowledge about items extracted from LLMs, which are transformed into task-compatible representations. \textbf{RLMRec}~\cite{ren2024representation} leverages LLMs to learn semantic user and item representations from textual signals and aligns them with collaborative relational information through cross-view representation learning. \textbf{PALR}~\cite{yang2023palr} fine-tunes a large language model as a ranking component that selects preferred items from retrieved candidates expressed in natural language. \textbf{LG (LettinGo)}~\cite{wang2025letting} explores diverse user profile candidates with LLMs and aligns profile generation with downstream recommendation performance via preference optimization. \textbf{Reason4Rec}~\cite{fang2025reason4rec} introduces a deliberative recommendation framework that incorporates explicit step-wise reasoning over user preferences to guide rating prediction.

\begin{table*}[t]
\scriptsize
\captionsetup{skip=5pt}

\centering
\setlength{\tabcolsep}{5pt} 
\renewcommand{\arraystretch}{1.2} 

\newcolumntype{Y}{>{\centering\arraybackslash}X}

\begin{tabularx}{\textwidth}{l *{12}{Y}}
\toprule
\textbf{Method} 
& \multicolumn{4}{c}{\textbf{Yelp}} 
& \multicolumn{4}{c}{\textbf{Amazon Music}} 
& \multicolumn{4}{c}{\textbf{Amazon Books}} \\
\cmidrule(lr){2-5} \cmidrule(lr){6-9} \cmidrule(lr){10-13}

& \textbf{MAE} & \textbf{RMSE} & \textbf{Acc (\%)} & \textbf{F1 (\%)} 
& \textbf{MAE} & \textbf{RMSE} & \textbf{Acc (\%)} & \textbf{F1 (\%)} 
& \textbf{MAE} & \textbf{RMSE} & \textbf{Acc (\%)} & \textbf{F1 (\%)} \\
\midrule

\multicolumn{13}{c}{\textbf{Qwen 3 (8B)}} \\
\midrule
10H & 1.1235 & 1.9478 & 23.17 & 27.54 & 0.9102 & 1.4021 & 39.26 & 46.58 & 0.9314 & 1.4527 & 37.63 & 45.19 \\
KAR~\cite{xi2024towards} & 0.7396 & 1.2184 & 55.34 & 48.67 & 0.7483 & 1.1380 & 58.65 & 60.29 & 0.7098 & 1.0923 & 56.17 & 58.78 \\
RLMRec~\cite{ren2024representation} & 0.8197 & 1.3312 & 47.15 & 42.46 & 0.7438 & 1.1069 & 54.89 & 57.65 & 0.7812 & 1.1584 & 52.86 & 55.93 \\
PALR~\cite{yang2023palr} & 0.7994 & 1.2876 & 48.53 & 43.19 & 0.6075 & 0.9531 & 57.35 & 56.77 & 0.7485 & 1.1187 & 54.24 & 56.38 \\
LG~\cite{wang2025letting} & 0.6632 & 1.1047 & 56.18 & 48.95 & 0.4737 & 0.8834 & 62.37 & 57.09 & 0.5821 & 0.9416 & 59.35 & 60.57 \\
R4Rec~\cite{fang2025reason4rec} & 0.7028 & 1.1523 & 55.69 & 47.73 & 0.5654 & 0.9635 & 58.69 & 54.67 & 0.6397 & 1.0098 & 58.47 & 56.84 \\
\textbf{Ours} & \textbf{0.5126} & \textbf{0.9485} & \textbf{61.23} & \textbf{55.18} & \textbf{0.3937} & \textbf{0.7564} & \textbf{67.96} & \textbf{63.89} & \textbf{0.4612} & \textbf{0.9089} & \textbf{64.38} & \textbf{59.27} \\
\midrule

\multicolumn{13}{c}{\textbf{LLaMA 3 (8B)}} \\
\midrule
10H & 1.0864 & 1.9532 & 22.09 & 27.30  & 0.7917 & 1.3346 & 38.13 & 46.87 & 0.8064 & 1.3866 & 37.15 & 45.27 \\
KAR~\cite{xi2024towards} & 0.6427 & 1.1668 & 54.51 & 47.98 & 0.5726 & 0.9033 & 57.53 & 59.92 & 0.5892 & 0.9614 & 55.87 & 58.21 \\
RLMRec~\cite{ren2024representation} & 0.7428 & 1.3572 & 46.74 & 42.11 & 0.6076 & 0.9886 & 53.78 & 57.42 & 0.6226 & 0.9477 & 52.12 & 55.79 \\
PALR~\cite{yang2023palr} & 0.7238 & 1.3265 & 47.72 & 43.29 & 0.5823 & 0.9222 & 56.73 & 59.31 & 0.5977 & 0.8855 & 55.06 & 57.62 \\
LG~\cite{wang2025letting} & 0.6196 & 1.1289 & 56.03 & 51.24 & 0.5204 & 0.9369 & 61.92 & 59.50 & 0.5543 & 0.7967 & 58.95 & 60.39 \\
R4Rec~\cite{fang2025reason4rec} & 0.7586 & 1.0418 & 55.80 & 53.00  & 0.5442 & 0.7722 & 60.86 & 54.88 & 0.6029 & 0.8345 & 59.70 & 56.35 \\
\textbf{Ours} & \textbf{0.5367} & \textbf{0.9687} & \textbf{60.87} & \textbf{54.74} & \textbf{0.4680} & \textbf{0.8277} & \textbf{63.30} & \textbf{60.60} & \textbf{0.5092} & \textbf{0.9500} & \textbf{63.42} & \textbf{58.12} \\
\bottomrule
\end{tabularx}
\caption{Performance on three datasets using Qwen 3 (8B) and LLaMA 3 (8B).}
\label{table:main_result}
\end{table*}

\begin{table*}[t!]
\scriptsize
\captionsetup{skip=5pt}

\centering
\setlength{\tabcolsep}{6pt}
\renewcommand{\arraystretch}{1.2}

\newcolumntype{Y}{>{\centering\arraybackslash}X}

\begin{tabularx}{\textwidth}{l *{9}{Y}}
\toprule
\textbf{Method} 
& \multicolumn{3}{c}{\textbf{Yelp}} 
& \multicolumn{3}{c}{\textbf{Amazon Music}} 
& \multicolumn{3}{c}{\textbf{Amazon Books}} \\
\cmidrule(lr){2-4} \cmidrule(lr){5-7} \cmidrule(lr){8-10}

& \textbf{NDCG@1} & \textbf{NDCG@5} & \textbf{NDCG@10}
& \textbf{NDCG@1} & \textbf{NDCG@5} & \textbf{NDCG@10}
& \textbf{NDCG@1} & \textbf{NDCG@5} & \textbf{NDCG@10} \\
\midrule

10H 
& 0.1823 & 0.2815 & 0.4928
& 0.1875 & 0.3796 & 0.5153
& 0.1841 & 0.3146 & 0.4263 \\

KAR\cite{xi2024towards}
& 0.2156 & 0.3298 & 0.5412
& 0.3018 & 0.4896 & 0.6015
& 0.2965 & 0.4715 & 0.5834 \\

RLMRec\cite{ren2024representation}
& 0.2419 & 0.3472 & 0.5587
& 0.3371 & 0.5434 & 0.6162
& 0.2748 & 0.4526 & 0.5719 \\

PALR\cite{yang2023palr}
& 0.2494 & 0.3563 & 0.5691
& 0.3395 & 0.5247 & 0.6115
& 0.2627 & 0.4634 & 0.5538 \\

LG\cite{wang2025letting}
& 0.3187 & 0.4685 & 0.5814
& 0.4012 & 0.5674 & 0.6489
& 0.3795 & 0.5189 & 0.6284 \\

R4Rec\cite{fang2025reason4rec}
& 0.2575 & 0.3792 & 0.5526
& 0.2928 & 0.5912 & 0.6343
& 0.3013 & 0.4928 & 0.5959 \\

\textbf{Ours}
& \textbf{0.3390} & \textbf{0.4873} & \textbf{0.6008}
& \textbf{0.5123} & \textbf{0.6165} & \textbf{0.7025}
& \textbf{0.4288} & \textbf{0.5638} & \textbf{0.6599} \\

\bottomrule
\end{tabularx}

\caption{Ranking performance under EASE-based \cite{steck2019ease} hard negatives.}

\label{tab:ease_ranking}
\end{table*}

\subsection{Main Results}
Table~\ref{table:main_result} presents a comparison of our proposed method against five baselines on three datasets: Amazon Music, Amazon Books, and Yelp. We use Qwen3-8B \citep{qwen3technicalreport} and LLaMA3-8B \citep{dubey2024llama} as both the profile generator and the downstream recommendation model, with a prediction temperature of 0. In addition to rating prediction results, we also report ranking-based evaluation results in Table~\ref{table:ranking_result}, which assess the effectiveness of the learned profiles from a downstream ranking perspective.

\textbf{Overall superiority over strong baselines.}  
As shown in Table~\ref{table:main_result}, our method consistently outperforms the strongest baselines across all three datasets under both Qwen3-8B and LLaMA3-8B. Under Qwen3-8B, our method achieves an accuracy of 61.23\% on Yelp, 67.96\% on Amazon Music and 64.38\% on Amazon Books, surpassing \textbf{LG (LettinGo)}~\cite{wang2025letting} by 5.05\%, 5.59\% and 5.03\%, respectively. Similar improvements are observed under LLaMA3-8B, indicating that the gains are stable across backbone models.

\textbf{Advantages over fixed-structure and deliberative baselines.}  
Compared with strong fixed-structure or deliberative methods such as \textbf{KAR}~\cite{xi2024towards} and \textbf{Reason4Rec}~\cite{fang2025reason4rec}, our approach consistently achieves lower prediction error and higher accuracy. For example, on Amazon Books with Qwen3-8B, our method reduces MAE to 0.4612, compared to 0.7098 for KAR and 0.6397 for Reason4Rec, demonstrating more effective abstraction of user preferences without relying on predefined reasoning templates.

\textbf{Consistent improvements in downstream ranking.}  
Table~\ref{table:ranking_result} reports the ranking results under the Qwen3-8B backbone. Our method consistently achieves the best performance across all three datasets and evaluation cutoffs. On Yelp, our approach reaches an NDCG@1 of 0.5619 and an NDCG@10 of 0.7443, clearly outperforming the strongest baseline methods. Similar trends are observed on Amazon Music, where our method attains 0.5347 at NDCG@1 and 0.7331 at NDCG@10, as well as on Amazon Books with NDCG@1 of 0.4866 and NDCG@10 of 0.7107. These results indicate that the learned profiles enable more accurate identification and ordering of relevant items within candidate sets, leading to superior ranking quality in downstream recommendation. We further evaluate ranking performance under a more challenging setting using EASE-based~\cite{steck2019ease} hard negatives, and observe consistent improvements; detailed results are provided in Appendix~\ref{app:ease_ranking}.

\subsection{Ablation Study}

\begin{table*}[t]
\small
\captionsetup{skip=5pt}

\centering
\setlength{\tabcolsep}{4pt}
\renewcommand{\arraystretch}{1.25}

\newcolumntype{M}{>{\raggedright\arraybackslash}p{3.8cm}}
\newcolumntype{Y}{>{\centering\arraybackslash}X}

\begin{tabularx}{\textwidth}{M *{12}{Y}}
\toprule
\textbf{Method} 
& \multicolumn{4}{c}{\textbf{Yelp}} 
& \multicolumn{4}{c}{\textbf{Amazon Music}} 
& \multicolumn{4}{c}{\textbf{Amazon Books}} \\
\cmidrule(lr){2-5} \cmidrule(lr){6-9} \cmidrule(lr){10-13}

& MAE & RMSE & Acc & F1
& MAE & RMSE & Acc & F1
& MAE & RMSE & Acc & F1 \\
\midrule

10H (History Only)
& 1.1235 & 1.9478 & 23.17 & 27.54
& 0.9102 & 1.4021 & 39.26 & 46.58
& 0.9314 & 1.4527 & 37.63 & 45.19 \\

Profile
& 0.7218 & 1.1863 & 55.48 & 48.09
& 0.6597 & 1.0218 & 58.67 & 57.48
& 0.6764 & 1.0469 & 57.14 & 57.68 \\

Profile + Cue\&Strategy
& 0.7085 & 1.1654 & 55.83 & 48.54
& 0.5708 & 0.9897 & 58.91 & 55.53
& 0.6389 & 1.0108 & 58.43 & 56.88 \\

\midrule

Profile + Joint Opt.(LG~\cite{wang2025letting})
& 0.6632 & 1.1047 & 56.18 & 48.95
& 0.4737 & 0.8834 & 62.37 & 57.09
& 0.5821 & 0.9416 & 59.35 & 60.57 \\

Profile + Cue\&Strategy + Joint Opt. (\textbf{Ours})
& \textbf{0.5126} & \textbf{0.9485} & \textbf{61.23} & \textbf{55.18}
& \textbf{0.3937} & \textbf{0.7564} & \textbf{67.96} & \textbf{63.89}
& \textbf{0.4612} & \textbf{0.9089} & \textbf{64.38} & \textbf{59.27} \\
\bottomrule
\end{tabularx}

\caption{Ablation study on different design configurations in \ourmethod using Qwen 3 (8B).}

\vspace{1mm}
\label{table:ablation}
\end{table*}

\begin{table*}[t!]
    \small  
    \captionsetup{skip=5pt}

    \centering
    \setlength{\tabcolsep}{3.5pt}  
    \renewcommand{\arraystretch}{1.2}  

    \newcolumntype{Y}{>{\centering\arraybackslash}X}

    \begin{tabularx}{\textwidth}{l *{12}{Y}}
        \toprule
        \textbf{Method} 
        & \multicolumn{4}{c}{\textbf{Yelp}} 
        & \multicolumn{4}{c}{\textbf{Amazon Music}} 
        & \multicolumn{4}{c}{\textbf{Amazon Books}} \\
        \cmidrule(lr){2-5} \cmidrule(lr){6-9} \cmidrule(lr){10-13}
        
        & \textbf{MAE} & \textbf{RMSE} & \textbf{Acc (\%)} & \textbf{F1 (\%)} 
        & \textbf{MAE} & \textbf{RMSE} & \textbf{Acc (\%)} & \textbf{F1 (\%)} 
        & \textbf{MAE} & \textbf{RMSE} & \textbf{Acc (\%)} & \textbf{F1 (\%)} \\
        \midrule
        10H+30P & 0.5126 & 0.9485 & 61.23 & 55.18 & \textbf{0.3883} & \textbf{0.7494} & \textbf{67.96} & \textbf{63.89} & 0.4612 & 0.9089 & \textbf{65.13} & \textbf{59.97} \\
        10H+50P & \textbf{0.4909} & \textbf{0.9207} & \textbf{62.43} & \textbf{56.24} & 0.3924 & 0.7543 & 67.88 & 63.87 & \textbf{0.4553} & \textbf{0.9023} & 64.62 & 59.52 \\
        10H+70P & 0.4987 & 0.9326 & 61.98 & 55.81 & 0.3937 & 0.7564 & 68.22 & 64.12 & 0.4608 & 0.9068 & 64.38 & 59.27 \\
        \bottomrule
    \end{tabularx}
    \caption{Impact of historical interaction length on profile quality (using Qwen 3 (8B)).}
    \label{table:history_length}
\end{table*}
\noindent \textit{\textbf{Effectiveness of profile generation, cue\&strategy, and user--item joint optimization.}}
Table~\ref{table:ablation} reports the ablation results under different design configurations. Starting from the history-only baseline (\textbf{10H}), introducing explicit \textbf{profile generation} yields substantial performance improvements across all datasets. For example, on Yelp, MAE is reduced from 1.1235 to 0.7218 and accuracy increases from 23.17\% to 55.48\%, confirming that textual profiles provide significantly more informative representations than raw interaction histories.

Adding the \textbf{cue\&strategy layer} on top of profile generation leads to modest but consistent gains. On Amazon Music, MAE further decreases from 0.6597 to 0.5708, while accuracy slightly improves from 58.67\% to 58.91\%. Although the numerical improvements introduced by strategy alone are limited, these results suggest that strategy discovery primarily reshapes how preference information is abstracted and expressed, rather than directly optimizing prediction accuracy in isolation.

When enabling \textbf{joint optimization} without strategy (i.e., the LettinGo~\cite{wang2025letting}-style configuration), performance improves more noticeably across datasets. On Amazon Music, accuracy increases to 62.37\%, compared to 58.67\% with profile generation alone. In our implementation, this setting corresponds to a reproduction of LettinGo, where profile generation and joint optimization are applied without the strategy layer. While the original method focuses primarily on user profiling, we extend the optimization to both user and item profiles to ensure a fair comparison.

The strongest and most consistent performance is achieved when \textbf{cue\&strategy discovery} and \textbf{joint optimization are combined}. Under this full configuration, accuracy reaches 61.23\% on Yelp, 67.96\% on Amazon Music, and 64.38\% on Amazon Books, with corresponding MAE values of 0.5126, 0.3937, and 0.4612, respectively. Compared with the LettinGo-style configuration, this setting yields consistent improvements across all datasets.
Overall, the ablation results indicate that while \textbf{profile generation} and \textbf{joint optimization} contribute substantially to performance gains, integrating \textbf{cue\&strategy discovery} further enhances the effectiveness of joint user--item optimization by providing a structured space for exploration and refinement.

\noindent \textit{\textbf{Impact of historical interaction length on profile quality.}}
Table~\ref{table:history_length} analyzes the impact of historical interaction length on profile quality. Across the three datasets, varying the number of historical interactions leads to only moderate performance differences, indicating that our method does not strongly depend on long histories. In particular, using a moderate history length (e.g., 30–50 interactions) already achieves competitive or best performance on most metrics, while further increasing the history length provides limited additional gains.

On Yelp and Amazon Books, extending the history length beyond this range does not consistently improve accuracy or F1 score and may slightly degrade performance, suggesting that excessive historical interactions can introduce noisy or less relevant signals. In contrast, Amazon Music exhibits relatively stable performance across different history lengths, indicating that user preferences in this domain are less sensitive to history truncation.

\begin{table*}[h!]
    \small
    \captionsetup{skip=5pt}
    \centering
    \setlength{\tabcolsep}{3.5pt}
    \renewcommand{\arraystretch}{1.2}

    \newcolumntype{Y}{>{\centering\arraybackslash}X}

    \begin{tabularx}{\textwidth}{l *{12}{Y}}
        \toprule
        \textbf{Setting} 
        & \multicolumn{4}{c}{\textbf{Yelp}} 
        & \multicolumn{4}{c}{\textbf{Amazon Music}} 
        & \multicolumn{4}{c}{\textbf{Amazon Books}} \\
        \cmidrule(lr){2-5} \cmidrule(lr){6-9} \cmidrule(lr){10-13}
        
        & \textbf{MAE} & \textbf{RMSE} & \textbf{Acc (\%)} & \textbf{F1 (\%)} 
        & \textbf{MAE} & \textbf{RMSE} & \textbf{Acc (\%)} & \textbf{F1 (\%)} 
        & \textbf{MAE} & \textbf{RMSE} & \textbf{Acc (\%)} & \textbf{F1 (\%)} \\
        \midrule
        DUET w/o RL 
        & 0.8283 & 1.3893 & 48.53 & 41.15 
        & 0.7322 & 1.1430 & 57.18 & 52.48 
        & 0.8741 & 1.4760 & 51.83 & 50.35 \\

        DUET (full) 
        & \textbf{0.5126} & \textbf{0.9485} & \textbf{61.23} & \textbf{55.18} 
        & \textbf{0.3937} & \textbf{0.7564} & \textbf{67.96} & \textbf{63.89} 
        & \textbf{0.4612} & \textbf{0.9089} & \textbf{64.38} & \textbf{59.27} \\
        \bottomrule
    \end{tabularx}
    \caption{Effect of RL-based optimization (using Qwen 3 (8B)). }
    \label{tab:rl_ablation}
\end{table*}
\noindent \textit{\textbf{Necessity of RL-based Optimization.}}
We compare the full model with a variant without RL (\textsc{\ourmethod w/o RL}) to isolate the effect of RL. As shown in Table~\ref{tab:rl_ablation}, removing RL leads to substantial performance degradation across all datasets (e.g., Yelp Accuracy drops from 61.23\% to 48.53\%), indicating that the gains cannot be attributed to prompt design alone.

Without RL, the generator reduces to a static mapping from interaction history to textual profiles, lacking adaptive selection of relevant signals. In contrast, RL enables optimizing profile construction under reward feedback, resulting in more discriminative representations.These results demonstrate that RL-based optimization is essential for \ourmethod.

\begin{figure*}[!t]
    \centering
\includegraphics[width=\textwidth]{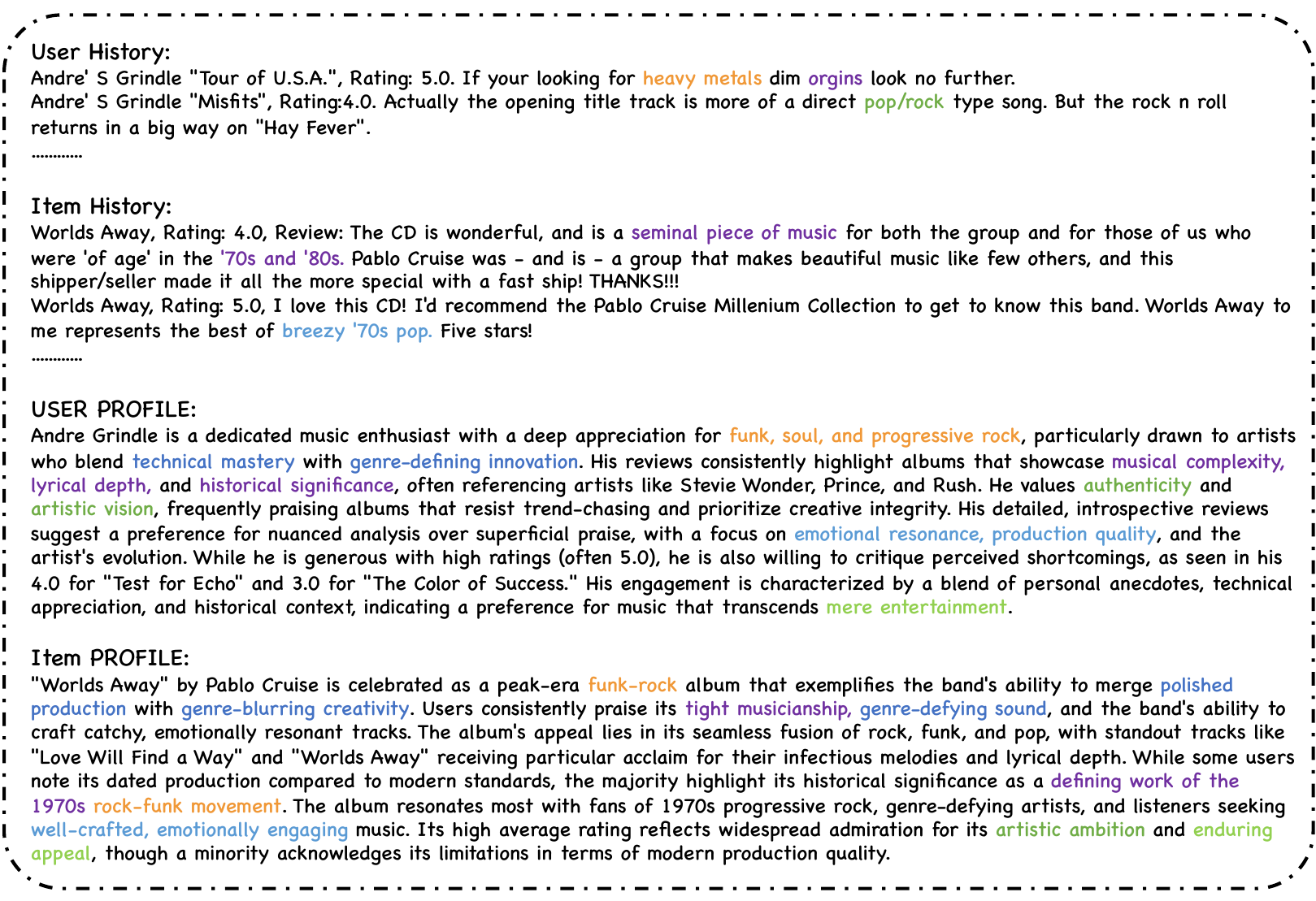}
    \caption{
    Illustration of the mutual correspondence between user and item.
    The highlighted regions demonstrate that user preferences summarized in the user profile align with the key attributes extracted in the item profile, which provides complementary information beyond raw histories and thus improves prediction accuracy.
    }
    \label{fig:user_item_case}
\end{figure*}

\subsection{Semantic Analysis of Generated Profiles}

To better understand the source of performance gains, we analyze whether the generated profiles exhibit meaningful semantic properties rather than serving as intermediate textual artifacts. We introduce two complementary metrics to characterize semantic compatibility and grounding.

\textbf{Semantic Alignment.}
We measure the embedding-level similarity between generated user and item profiles using \texttt{all-mpnet-base-v2} from Sentence-Transformers~\cite{reimers2019sentence}. For each user--item pair, we compute cosine similarity:
\begin{equation}
\mathrm{Align}(u,i) = \frac{\mathbf{e}_u \cdot \mathbf{e}_i}{\|\mathbf{e}_u\| \|\mathbf{e}_i\|}
\end{equation}
where $\mathbf{e}_u$ and $\mathbf{e}_i$ denote the embedding vectors of the generated user and item profiles. Higher values indicate stronger semantic compatibility between modeled user preferences and item characteristics.

\textbf{Coverage (Faithfulness).}
We measure token-level grounding as:
\begin{equation}
\mathrm{Cov} = \frac{|\mathrm{Tokens}(\mathrm{profile}) \cap \mathrm{Tokens}(\mathrm{history})|}{|\mathrm{Tokens}(\mathrm{profile})|}
\end{equation}
This metric quantifies how much of the generated profile is supported by historical textual evidence. We report coverage separately for user and item profiles.

As shown in Table~\ref{tab:alignment_coverage}, \ourmethod achieves the highest semantic alignment across all datasets while maintaining comparable coverage. Compared to existing methods, which either rely on extractive compression (e.g., KAR~\cite{xi2024towards}) or generate free-form descriptions with weaker grounding (e.g., RLMRec~\cite{ren2024representation} and PALR~\cite{yang2023palr}), \ourmethod consistently attains stronger alignment without sacrificing coverage. For instance, LG~\cite{wang2025letting} achieves relatively high alignment but exhibits less stable grounding across datasets, while RLMRec attains higher coverage at the cost of weaker semantic alignment. In contrast, \ourmethod maintains both high alignment and mid-to-high coverage, suggesting a more effective balance between semantic abstraction and evidence preservation. 

\begin{table*}[t!]
\scriptsize
\captionsetup{skip=5pt}

\centering
\setlength{\tabcolsep}{6pt}
\renewcommand{\arraystretch}{1.2}

\newcolumntype{Y}{>{\centering\arraybackslash}X}

\begin{tabularx}{\textwidth}{l *{9}{Y}}
\toprule
\textbf{Method} 
& \multicolumn{3}{c}{\textbf{Yelp}} 
& \multicolumn{3}{c}{\textbf{Amazon Music}} 
& \multicolumn{3}{c}{\textbf{Amazon Books}} \\
\cmidrule(lr){2-4} \cmidrule(lr){5-7} \cmidrule(lr){8-10}

& \textbf{Align} & \textbf{User Cov.} & \textbf{Item Cov.}
& \textbf{Align} & \textbf{User Cov.} & \textbf{Item Cov.}
& \textbf{Align} & \textbf{User Cov.} & \textbf{Item Cov.} \\
\midrule

10H 
& 0.3902 & /(1.00) & /(1.00)
& 0.3577 & /(1.00) & /(1.00)
& 0.2604 & /(1.00) & /(1.00) \\

KAR\cite{xi2024towards}
& 0.4932 & 0.1745 & 0.1823
& 0.4807 & 0.2170 & 0.2391
& 0.5702 & 0.1877 & 0.1544 \\

RLMRec\cite{ren2024representation}
& 0.4010 & 0.2646 & \textbf{0.3773}
& 0.3931 & 0.3526 & \textbf{0.4889}
& 0.4436 & 0.2994 & \textbf{0.3506} \\

PALR\cite{yang2023palr}
& 0.4715 & \textbf{0.3675} & 0.1917
& 0.4938 & 0.2197 & 0.2390
& 0.5216 & 0.2050 & 0.2267 \\

LG\cite{wang2025letting}
& 0.5709 & 0.2378 & 0.2682
& 0.5109 & 0.2546 & 0.3749
&  0.5506 & \textbf{0.3883} & 0.3364 \\

R4Rec\cite{fang2025reason4rec}
& 0.4882 & 0.2348 & 0.2330
& 0.4208 & 0.3328 & 0.3595
& 0.4946 & 0.2831 & 0.2949 \\

\textbf{Ours}
& \textbf{0.6382} & 0.2880 & 0.3429
& \textbf{0.5947} & \textbf{0.4002} & 0.4482
& \textbf{0.7287} & 0.3457 & 0.3127 \\

\bottomrule
\end{tabularx}

\caption{Semantic alignment and coverage of generated profiles. Cov. denotes the fraction of tokens grounded in historical text.}

\label{tab:alignment_coverage}
\end{table*}

\subsection{Case Study}
Figure~\ref{fig:user_item_case} presents a representative case study that demonstrates how semantically aligned user and item profiles enable accurate prediction that transcends the limitations of sparse raw interaction histories. In the user profile, initially scattered and fragmented preference cues are systematically distilled into a stable and coherent preference structure favoring \textit{funk, soul, and progressive rock} (orange), coupled with a pronounced emphasis on \textit{musical complexity} and \textit{historical significance} (purple). The corresponding item profile exhibits striking symmetry, characterizing the album as an exemplary \textit{funk-rock} composition (orange) and positioning it as a \textit{defining and culturally influential release of the 1970s} (purple). This example shows
519 that the learned profiles capture meaningful preference–attribute correspondence that is difficult to recover from individual reviews alone.
\section{Conclusion}
\vspace{-2pt}
In this paper, we propose \ourmethod, a closed-loop framework for jointly generating user and item textual profiles for recommendation. Unlike prior methods that rely on fixed templates or independently constructed profiles, DUET treats profile generation as an exploration problem and aligns representations directly with downstream recommendation performance.

Specifically, \ourmethod integrates cue-based initialization, adaptive strategy construction, and feedback-driven joint optimization to produce flexible yet task-aligned user–item profiles. By optimizing both profiles in a shared semantic space, the framework reduces semantic mismatch and captures interaction-relevant signals that are difficult to recover from raw histories or static prompts alone.

Experiments on multiple real-world datasets demonstrate that \ourmethod consistently outperforms strong baselines under different backbone models, validating the effectiveness of joint profiling and reinforcement learning–based optimization. These results suggest that adaptive, interaction-aware textual profiles provide a promising direction for more effective interpretable and performance-oriented recommendation systems.

\newpage
\section*{Limitations}

Despite its effectiveness, our approach has several limitations. First, the proposed framework relies on large language models for both profile generation and downstream recommendation, which introduces additional computational overhead during training and inference. While our experiments show consistent gains across different backbones, the overall efficiency may be constrained when scaling to very large user or item sets.
Second, the quality of the generated profiles is inherently dependent on the underlying LLMs and prompting strategies. Variations in model capacity or prompt sensitivity may lead to differences in profile stability, which we do not explicitly control in the current design.
Finally, our evaluation focuses on text-rich recommendation scenarios where sufficient historical reviews are available. The effectiveness of the proposed strategy in domains with extremely sparse textual signals or in non-textual modalities remains to be further explored.

\section*{Data, Privacy, and Ethics Considerations}
All datasets used in this work are publicly released by their original owners, and we follow the official terms of use associated with each dataset, utilizing the data solely for research purposes as permitted by their respective licenses; for reproducibility, the official dataset sources are cited in the corresponding sections of this paper, and we do not redistribute any raw user-generated content such as review text as part of this work. This research does not attempt to identify or infer the identity of any individual user, and prior to training and downstream profile generation, we apply text filtering procedures to remove direct personal identifiers where applicable; the models are trained solely on publicly available data, and no additional private or proprietary user information is introduced. Furthermore, we do not release generated user profiles or any per-user outputs produced during the modeling process, and all reported results are presented in aggregate form for evaluation purposes only. From a security perspective, the associated code repository has been scanned for potential secrets, and any internal endpoints or confidential configuration information have been removed prior to release.
\bibliography{custom}

\appendix
\newpage
\section{Experiment Setup}
\subsection{Data Curation}

\begin{table}[ht]
\centering
\small
\setlength{\tabcolsep}{4pt}
\caption{Statistical details of the evaluation datasets.}
\label{tab:dataset_stats}
\begin{tabular}{lcccccc}
\toprule
\textbf{Dataset} & \textbf{\#Train} & \textbf{\#Valid} & \textbf{\#Test} & \textbf{\#User} & \textbf{\#Item} \\
\midrule
Music & 43,071 & 3,271 & 1,296 & 4,183 & 2,660 \\
Book  & 71,972 & 6,144 & 5,541 & 13,863 & 13,515 \\
Yelp  & 51,497 & 4,757 & 4,328 & 8,453 & 13,426 \\
\bottomrule
\end{tabular}
\end{table}

We conduct experiments on three widely used real-world datasets:
\begin{itemize}[leftmargin=*]
\item \textbf{Amazon Music (Music)}: This refers to the ``Digital Music'' subset of the well-known Amazon Product dataset\footnote{\url{https://cseweb.ucsd.edu/~jmcauley/datasets/amazon/links.html}.}, which records rich user reviews, ratings, and textual information about items, such as titles, across a broad range of product categories, on the Amazon platform.2

\item \textbf{Amazon Book (Book)}: This refers to the ``Book'' subset of the Amazon Product dataset.
\item \textbf{Yelp:}
This refers to the Yelp Open dataset\footnote{\url{https://business.yelp.com/data/resources/open-dataset/}.}, which includes user reviews, ratings for businesses such as restaurants and retail shops, as well as textual information about the businesses. It is widely used in recommendation tasks~\cite{rating_predict1}.
\end{itemize}

We use the entire Music dataset for experiments, while for the Book and Yelp datasets, we utilize only a subset due to their large size. For the Book dataset, we use data from the last two months, and for the Yelp dataset, we use data from the last six months. For each dataset, we split it into training, validation, and test sets based on the timestamps of interactions, ensuring that test interactions occur after all training and validation interactions to prevent information leakage~\cite{Ji2023DataLeakage}.

Regarding data filtering, following prior work~\cite{liu2019nrpa}, we adopt a 5-core setting to filter the data and exclude cold-start users and items—those not appearing in the training set—from the validation and test sets. The statistical details of the processed dataset are provided in Table~\ref{tab:dataset_stats}.

\subsubsection{Implementation Details}
In our experiments, we primarily employ Qwen3-8B~\cite{qwen3technicalreport} and LLaMA3 8B Instruct ~\cite{dubey2024llama,touvron2023llama} as
both the recommendation model and the profile generation model. The training process is implemented using the TRL~\cite{von_Werra_TRL_Transformer_Reinforcement}. Key hyperparameters, such as batch size and learning rate, are determined through grid search to achieve optimal performance. More details can be found in our code.

\subsection{Baseline Prompts}\label{appendix:baselineP}

\begin{promptbox}[title={KAR Prompt}]
\textbf{Task:} Analyze user preferences based on business reviewing history \\
\textbf{Input:} \{user\_history\} - User's business reviewing history with sentiments over time \\
\textbf{Instructions:} \\
1. Analyze the user's preferences considering business names and categories \\
2. Take into account sentiment patterns over time \\
3. Provide clear explanations based on reviewing history details \\
4. Consider other pertinent factors that may influence preferences
\end{promptbox}

\begin{promptbox}[title={PALR Prompt}]
\textbf{Task:} Summarize user preferences using keywords.

\textbf{Input:} \{user\_history\} - historical businesses with user sentiments.

\textbf{Output Format:} An itemized list ranked by importance.

\textbf{Template:}
\begin{itemize}
  \item KEY\_WORD\_1: "HISTORY\_BUSINESS\_1", "HISTORY\_BUSINESS\_2"
  \item KEY\_WORD\_2: "HISTORY\_BUSINESS\_3"
\end{itemize}

\textbf{Instructions:}
\begin{enumerate}
  \item Extract key preference indicators from user interaction history.
  \item Rank keywords by importance.
\end{enumerate}
\end{promptbox}

\begin{promptbox}[title={RLMRec Prompt}]
\textbf{Role:} Business recommendation assistant \\
\textbf{Task:} Determine business types a user is likely to enjoy \\
\textbf{Input Format:} \\
• Title: Business name \\
• Categories: Business categories \\
• Sentiment: User sentiment toward business \\
\\
\textbf{Output Requirements:} \\
1. JSON format only \\
2. Structure: \\
\{\\
\phantom{\{}"summarization": "Types of businesses user likely enjoys" ($\leq$100 words),\\
\phantom{\{}"reasoning": "Brief explanation for summarization" (no word limit)\\
\}\\
3. No additional text outside JSON \\
\\
\textbf{Input:} INTERACTION ITEMS: \{user\_history\}
\end{promptbox}

\begin{promptbox}[title={LG Prompt}]
You will serve as an assistant to help me generate a user profile based on this user's sentiments history to better understand this users' interest and thus predict his/her sentiment about a target item. I will provide you with some behavior history of the user in this format: [item attributes and sentiment]. The user profile you generate should contain as much useful content as possible to help predict the user's sentiment towards a new business.\\
\textbf{USER HISTORY:} \{user\_history\}.\\
\textbf{PROFILE YOU GENERATE:}
\end{promptbox}

\begin{promptbox}[title={R4Rec Prompt (Reasoner)}]
\textbf{User Review History} \\
$\langle H_u$ organized as below$\rangle$ \\
1. Title of Item 1 \\
Positive Aspects: [Aspect 1], [Aspect 2], ... \\
Negative Aspects: [Aspect 1], [Aspect 2], ... \\
User Preference Elements: [Preference 1], [Preference 2], ... \\
2. Title of Item 2 \\
Positive Aspects: [Aspect 1], [Aspect 2], ... \\
Negative Aspects: [Aspect 1], [Aspect 2], ... \\
User Preference Elements: [Preference 1], [Preference 2], ... \\
... \\
\textbf{Item Review History by Other Users} \\
$\langle H_i$ organized in the same format as above$\rangle$ \\
Task: Analyze whether the user will like the new Music $i$ based on the user's preferences and the item's features. Provide your rationale in one concise paragraph.
\end{promptbox}

\subsection{Example of Cue}\label{app:cue}
\noindent The following examples illustrate how raw signals are distilled into cues. As shown on the top, user cues emphasize historical preferences, while item cues highlight metadata and user-group patterns. Together, they provide minimal but informative hypotheses for profile exploration.  

\begin{figure}[h]
\centering
\colorbox{gray!10}{%
\begin{minipage}{1\linewidth}
\textbf{Examples of User Cues}\\
\textcolor{black!10!blue}{\emph{“enjoys retro puzzle games”}} — derived from repeated engagement with classic titles.\\
\textcolor{black!10!blue}{\emph{“prefers concise product reviews”}} — inferred from a pattern of short, direct comments.\\
\textcolor{black!10!blue}{\emph{“tends to give high ratings but rarely comments”}} — highlighting consistency but limited feedback.
\end{minipage}}
\end{figure}
\vspace{-5mm}
\begin{figure}[h]
\centering
\colorbox{gray!20}{%
\begin{minipage}{\linewidth}
\textbf{Examples of Item Cues}\\
\textcolor{black!10!purple}{\emph{“lightweight trail-running shoes”}} — derived from product metadata.\\
\textcolor{black!10!purple}{\emph{“popular among budget-conscious users”}} — inferred from purchase patterns.\\
\textcolor{black!10!purple}{\emph{“stylized with retro aesthetics”}} — extracted from item descriptions.
\end{minipage}}
\end{figure}

\section{Additional Experiments}
\subsection{Non-LLM Baseline via Extractive Summarization}

To examine whether the gains of DUET stem from improved semantic representations rather than generic text generation, we introduce a non-LLM baseline based on extractive summarization. Specifically, we apply TextRank~\cite{mihalcea2004textrank} to select salient sentences from user histories and construct user profiles without using any generative model.

The extracted summaries are then fed into the same downstream predictor for rating estimation. This baseline isolates the effect of readable summarization from representation learning.

Table~\ref{tab:textrank} shows that TextRank improves over simple history truncation (10H), but remains significantly worse than DUET across all datasets. This indicates that coherent summaries alone are insufficient, and that the gains of DUET arise from learned semantic abstraction rather than extractive compression.

\begin{table*}[b]
\scriptsize
\captionsetup{skip=5pt}

\centering
\setlength{\tabcolsep}{6pt}
\renewcommand{\arraystretch}{1.2}

\newcolumntype{Y}{>{\centering\arraybackslash}X}

\begin{tabularx}{\textwidth}{l *{12}{Y}}
\toprule
\textbf{Method} 
& \multicolumn{4}{c}{\textbf{Yelp}} 
& \multicolumn{4}{c}{\textbf{Amazon Music}} 
& \multicolumn{4}{c}{\textbf{Amazon Books}} \\
\cmidrule(lr){2-5} \cmidrule(lr){6-9} \cmidrule(lr){10-13}

& \textbf{MAE} & \textbf{RMSE} & \textbf{Acc (\%)} & \textbf{F1 (\%)} 
& \textbf{MAE} & \textbf{RMSE} & \textbf{Acc (\%)} & \textbf{F1 (\%)} 
& \textbf{MAE} & \textbf{RMSE} & \textbf{Acc (\%)} & \textbf{F1 (\%)} \\
\midrule

10H 
& 0.8283 & 1.3893 & 48.53 & 41.15 
& 0.7322 & 1.1430 & 57.18 & 52.48 
& 0.8741 & 1.4760 & 51.83 & 50.35 \\

TextRank
& 0.8104 & 1.1323 & 50.51 & 30.23
& 0.5914 & 0.8223 & 61.19 & 29.68
& 0.6328 & 0.8854 & 59.59 & 29.48 \\

\textbf{Ours}
& \textbf{0.5126} & \textbf{0.9485} & \textbf{61.23} & \textbf{55.18}
& \textbf{0.3937} & \textbf{0.7564} & \textbf{67.96} & \textbf{63.89}
& \textbf{0.4612} & \textbf{0.9089} & \textbf{64.38} & \textbf{59.27} \\

\bottomrule
\end{tabularx}

\caption{Comparison with a non-LLM extractive summarization baseline (TextRank).}
\label{tab:textrank}
\end{table*}

\subsection{Ranking under Hard Negative Sampling}
\label{app:ease_ranking}

To construct a more challenging ranking scenario, we replace random negatives with hard negatives generated by a collaborative filtering model. Specifically, for each user, we retrieve high-scoring items from an EASE\cite{steck2019ease} model that the user has not interacted with, and combine them with the ground-truth item to form the candidate set.

Table~\ref{tab:ease_ranking} reports the results. Compared to random sampling, performance decreases for all methods due to increased difficulty, while DUET consistently maintains the best performance across datasets. This indicates that the improvements are robust and not limited to trivial ranking scenarios.

\begin{table*}[b]
\scriptsize
\captionsetup{skip=5pt}

\centering
\setlength{\tabcolsep}{6pt}
\renewcommand{\arraystretch}{1.2}

\newcolumntype{Y}{>{\centering\arraybackslash}X}

\begin{tabularx}{\textwidth}{l *{9}{Y}}
\toprule
\textbf{Method} 
& \multicolumn{3}{c}{\textbf{Yelp}} 
& \multicolumn{3}{c}{\textbf{Amazon Music}} 
& \multicolumn{3}{c}{\textbf{Amazon Books}} \\
\cmidrule(lr){2-4} \cmidrule(lr){5-7} \cmidrule(lr){8-10}

& \textbf{NDCG@1} & \textbf{NDCG@5} & \textbf{NDCG@10}
& \textbf{NDCG@1} & \textbf{NDCG@5} & \textbf{NDCG@10}
& \textbf{NDCG@1} & \textbf{NDCG@5} & \textbf{NDCG@10} \\
\midrule

10H 
& 0.1823 & 0.2815 & 0.4928
& 0.1875 & 0.3796 & 0.5153
& 0.1841 & 0.3146 & 0.4263 \\

KAR\cite{xi2024towards}
& 0.2156 & 0.3298 & 0.5412
& 0.3018 & 0.4896 & 0.6015
& 0.2965 & 0.4715 & 0.5834 \\

RLMRec\cite{ren2024representation}
& 0.2419 & 0.3472 & 0.5587
& 0.3371 & 0.5434 & 0.6162
& 0.2748 & 0.4526 & 0.5719 \\

PALR\cite{yang2023palr}
& 0.2494 & 0.3563 & 0.5691
& 0.3395 & 0.5247 & 0.6115
& 0.2627 & 0.4634 & 0.5538 \\

LG\cite{wang2025letting}
& 0.3187 & 0.4685 & 0.5814
& 0.4012 & 0.5674 & 0.6489
& 0.3795 & 0.5189 & 0.6284 \\

R4Rec\cite{fang2025reason4rec}
& 0.2575 & 0.3792 & 0.5526
& 0.2928 & 0.5912 & 0.6343
& 0.3013 & 0.4928 & 0.5959 \\

\textbf{Ours}
& \textbf{0.3390} & \textbf{0.4873} & \textbf{0.6008}
& \textbf{0.5123} & \textbf{0.6165} & \textbf{0.7025}
& \textbf{0.4288} & \textbf{0.5638} & \textbf{0.6599} \\

\bottomrule
\end{tabularx}

\caption{Ranking performance under EASE-based\cite{steck2019ease} hard negatives.}

\label{tab:ease_ranking}
\end{table*}

\subsection{Robustness under Preference Diversity}

We further analyze the robustness of DUET under varying levels of user preference diversity. We use the variance of historical ratings as a proxy for preference stability: low variance indicates consistent preferences, while high variance corresponds to diverse or potentially conflicting signals.

We partition users into three groups based on percentile thresholds (bottom 33\%, middle 33\%, top 33\% of rating variance) and evaluate performance within each group.

As shown in Table~\ref{tab:robustness}, performance degrades smoothly as preference diversity increases across all datasets. Importantly, the degradation is gradual rather than catastrophic, indicating that DUET remains stable under heterogeneous or noisy interaction histories.

\begin{table*}[tp]
\scriptsize
\captionsetup{skip=5pt}

\centering
\setlength{\tabcolsep}{5pt}
\renewcommand{\arraystretch}{1.2}

\newcolumntype{Y}{>{\centering\arraybackslash}X}

\begin{tabularx}{\textwidth}{l *{15}{Y}}
\toprule
\textbf{Variance Group} 
& \multicolumn{5}{c}{\textbf{Yelp}} 
& \multicolumn{5}{c}{\textbf{Amazon Music}} 
& \multicolumn{5}{c}{\textbf{Amazon Books}} \\
\cmidrule(lr){2-6} \cmidrule(lr){7-11} \cmidrule(lr){12-16}

& \textbf{\#Samp.} & \textbf{MAE} & \textbf{RMSE} & \textbf{Acc (\%)} & \textbf{F1 (\%)}
& \textbf{\#Samp.} & \textbf{MAE} & \textbf{RMSE} & \textbf{Acc (\%)} & \textbf{F1 (\%)}
& \textbf{\#Samp.} & \textbf{MAE} & \textbf{RMSE} & \textbf{Acc (\%)} & \textbf{F1 (\%)} \\
\midrule

Stable 
& 1530 & 0.4017 & 0.8664 & 71.76 & 68.96
& 432  & 0.2368 & 0.5990 & 82.56 & 86.89
& 1876 & 0.3171 & 0.8403 & 76.64 & 72.56 \\

Moderate 
& 1354 & 0.4773 & 0.8309 & 60.09 & 46.26
& 436  & 0.3921 & 0.6628 & 64.16 & 47.02
& 1827 & 0.4495 & 0.8059 & 66.73 & 55.45 \\

Diverse 
& 1444 & 0.6632 & 1.1198 & 51.13 & 45.95
& 428  & 0.5536 & 0.9607 & 57.10 & 59.27
& 1838 & 0.6199 & 1.0603 & 49.52 & 46.35 \\

\bottomrule
\end{tabularx}

\caption{Performance of DUET under different levels of user preference diversity (measured by rating variance). \#Samp. denotes the number of samples in each group.}

\label{tab:robustness}
\end{table*}
\section{The Use of Large Language Models}
We used a Large Language Model (LLM) only as a writing assistant to polish the language of the manuscript (\emph{e.g.}, grammar refinement, style adjustment, and clarity improvement). The research ideas, methodology design, experiments, and analysis were entirely conceived, implemented, and validated by the authors without reliance on the LLM. The LLM did not contribute to research ideation, experimental design, or result interpretation.

\end{document}